\documentclass[twocolumn,showpacs,preprintnumbers,amsmath,amssymb]{revtex4}

\usepackage[dvips]{graphicx}
\usepackage{epsfig}

\begin{document}

\preprint{ }

\title{ Equations of Motion with Multiple Proper Time: A New Interpretation of Basic Quantum Physics }  
\author{\normalsize Xiaodong Chen}
\email{xiaodong.chen@gmail.com} 

\date{\today}

\begin{abstract} 
  Equations of motion for single particle under two proper time model and three proper time model 
have been proposed and analyzed. 
The motions of particle are derived from pure classical method but they exhibit the same properties of quantum 
physics: the quantum wave equation, de Broglie equations, uncertainty relation, statistical result  
of quantum wave-function. This shows us a possible new way to interpret quantum physics. 
We will also prove that physics with multiple proper time does not cause causality problem. 
\end{abstract}

\maketitle

\section{Introduction} \label{INTRO}

From Thirring \cite{Thirring1} and Kaluza\cite{Kal21} 's 5-dimensional space-time to
today's superstring theory, Physics with extra space-time dimensions have been studied for about 90 years.
One of questions the theories have to face is: ``do we have any observation of extra
dimension in nature? '' For more than 90 years in another area, from Einstein to Bohm\cite{bohm}'s quantum hidden
variable theory, people have been looking for a way to find classical interpretation of 
quantum physics. The purpose of this paper is to show that it is possible to interpret quantum physics 
by adding extra time dimension (or dimensions) in classical physics.

 The extra space-time dimensions in most theories are space dimensions. People believed
that extra time dimension could cause causality problem \cite{Collins1}. I.Bars and etc \cite{IBars1}
\cite{IBars2}\cite{Romero} proposed two timelike dimensions in string theory which called 
``two time physics''. I.Bars' papers could raise questions like:
``Does two time-like dimensions necessarily mean two dimensional time? ";
``Does that create causality problem? '' One of the most important questions is: 
if there are two dimensional time, a particle should be able to move under two 
independent ``proper time'', how does the motion look like in basic physics?

In this paper, multiple proper times are introduced in classical physics. 
In addition, a particle is always treated as a point particle in this paper. Section 
\ref{TwoTime} will introduce the model of two independent proper time. 
The world lines equations and motion of equation of single particle under two proper 
time are derived. Section \ref{ThreeTime} will introduce the model of three dimensional 
time.
In section \ref{Interp}, the interpretation of quantum physics is discussed. In section
\ref{Causality}, Causality under multiple dimensional time is discussed. Section
\ref{Interp} will show that
multiple dimensional time will not cause causality problem.

\section{Two proper time model for classical single particle} \label{TwoTime}

Relativity introduced symmetry between time and space into physics. In Riemann Space, the only difference between
time and space are their signatures (+ and -);
but there is one asymmetrical properties which 
time is different from space -- proper time $\tau$. $\tau$ plays as a special affine parameter
in Relativity. In fact, $\tau$ is the one we called ``time''
in our common life(in low speed world). If there is n-dimensional time in the world, 
there should be n-dimensional proper time.  

In this section, we will study how the motion of single particle looks like if there are two proper
time.
Throughout this paper, $\tau$ indicates the first proper time which is the proper time in relativity. 
$\sigma$ indicates the second proper time. We call world line $\tau$ if particle moves along 
world line with proper time $\tau$, and world line $\sigma$ if particle moves along
world line with proper time $\sigma$, 
We made some reasonable assumptions to second proper time:

1)World lines $\sigma$ are orthogonal to world lines $\tau$.

2)$\sigma$ and $\tau$ are independent; i.e., when 
particle's position on world line $\tau$ is unchanged, the particle can still move on 
world line $\sigma$ and vice versa.

3)Similar to Kaluza \cite{Kal21} and Klein \cite{Kle26a}'s idea, we apply cylinder condition on
$\sigma$: $\sigma$ is a loop with value from 0 to $2\pi$.

For free particle with constant velocity, let $x_{\alpha\tau} (\alpha = 0..3)$ be 4-dimension coordinates on 
world line $\tau$; $u = \sqrt{u_i u^i}, (i=1,2,3) $ which 
is speed of particle on world line $\tau$;
$x_{\alpha\sigma} $ be 4-dimension coordinates on 
world line $\sigma$; $v = \sqrt{v_i v^i}, (i=1,2,3) $ which is speed of particle on world line 
$\sigma$. $t_\tau$ as time on
world line $\tau$, $t_\sigma$ as time on world line $\sigma$, then
\begin{eqnarray}
x_{0\tau} = c t_\tau \; \; \; \; x_{i\tau} = u_i t_\tau  \; \;, \;  \; 
t_\tau = \frac{\tau}{\sqrt{1-\frac{u^{2}}{c^2}}}  \;\; \;  \; 
\label{WorldLine1} 
\end{eqnarray}
where i=1,2,3; c is the speed of light. Momentum and energy can be defined as
\begin{eqnarray}
p_i = mu_i  \;\;, \; \;  E = mc^2
\end{eqnarray}
where i=1,2,3 ; m is mass of particle. Then from special relativity, world line $\tau$ satisfies condition 
\begin{equation}
\tau = \frac{i}{m_0}(Et - p_{1}x_{1} - p_{2}x_{2} - p_{3}x_{3})
\label{tau}
\end{equation}
where $m_0$ is static mass. Fig1 draws world line $\tau$ and world line 
$\sigma$ on $x_0-x_i$ plane in Minkowski Space.

\begin{figure}[t]
\begin{center}
\leavevmode
\hbox{%
\epsfxsize=2.5in
\epsffile{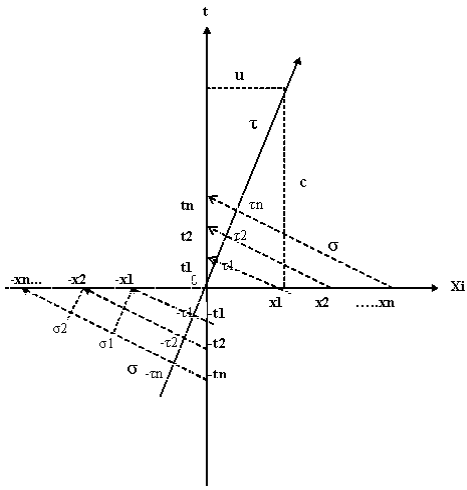}}
\caption{World line $\tau$ and world line $\sigma$ on $t-x_i$ plane in Minkowski space.
Particle can move along both world lines, slope of $\tau$ is u/c.
At $t=0$, the single particle will be shown at many positions:
$-x_1, -x_2, ... -x_n$ with different values of $\tau$ and $\sigma$ 
$(-\tau_1,\sigma_1)..(-\tau_n,\sigma_n)$. Also the single particle
will be shown at $x=0$ at different time: $t_1, t_2, .. t_n$ with different $\tau$ and $\sigma$
avalues; where
$x_n = h/mu$ and $t_n = h/mc^2$ which are de Broglie wavelength and period. }
\end{center}
\label{fig:Fig1}
\end{figure}

Slope of world line $\tau$ is  $x_{i\tau}/x_{0\tau}$,
Slope of world line $\sigma$ is  $x_{i\sigma}/x_{0 \sigma} $,
and $x_{i\tau}/x_{0\tau} = x_{0\sigma}/x_{i\sigma}$ since they are perpendicular
to each other,
so
\begin{equation}
v_i = \frac{c^2}{u_i} 
\label{Speed}
\end{equation}
$v_i > c$ and $\sigma$
is space like. 
$\tau$ and $\sigma$ are independent; On each point of world line $\tau$, particle can move
on world line $\sigma$. 

Fig2. draws world line $\tau$ and world line 
$\sigma$ on $x_0-x_i$ plane in Riemann Space.

\begin{figure}[t]
\begin{center}
\leavevmode
\hbox{%
\epsfxsize=2.5in
\epsffile{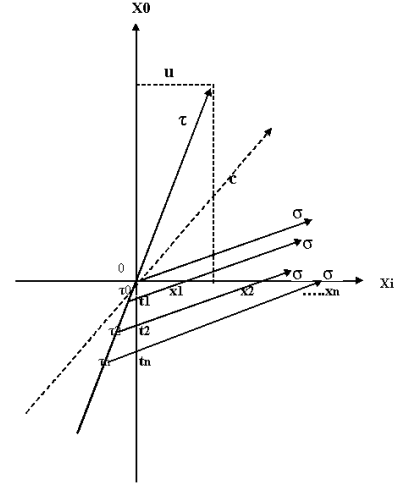}}
\caption{World line $\tau$ and world line $\sigma$ on $X_0-X_i$ plane in Riemann space.
At each points of world line $\tau$, the particle can move along world lines $\sigma$, 
slope of $\tau$ is u/c. Each world lines $\sigma$ parallel to each other with slope: 
$v/c = c/u$. In Riemann space, $\tau$ and $\sigma$ are orthogonal to each other.
At $t=0$, the single particle will be shown at many positions:
$x_1, x_2, ... x_n$ with different values of $\tau$ and $\sigma$ 
$(\tau_1,\sigma_1)..(\tau_n,\sigma_n)$. Also the single particle
will be shown at $x=0$ at different time: $t_1, t_2, .. t_n$ with
different values of $\tau, sigma$; where
$x_n = h/mu$ and $t_n = h/mc^2$ which are de Broglie wavelength 
and period. }
\end{center}
\label{fig:Fig2}
\end{figure}

The equations of motion of particle are:
\begin{eqnarray}
x_0(\tau, \sigma) = \frac{\tau}{\sqrt{1-\frac{u^{2}}{c^2}}} + i\frac{\sigma}{\sqrt{1-\frac{v^{2}}{c^2}}}
   + x_0(\tau_0,\sigma_0)  \\
x_i(\tau, \sigma) = \frac{u_i\tau}{\sqrt{1-\frac{u^{2}}{c^2}}} - i\frac{\sigma}{\sqrt{1-\frac{v^{2}}{c^2}}}
   + x_i(\tau_0,\sigma_0)  \; \; 
\label{MotionEq1}
\end{eqnarray}
where i=1,2,3 ; $x_0(\tau_0,\sigma_0)$ is initial value of $x_0$, $x_i(\tau_0,\sigma_0)$ is initial 
value of $x_i$; imaginary number i is to keep $\sigma$ and $x_\alpha$ to be real value
since $v >> c$. 
Under equation (\ref{MotionEq1}), particle's position is determined by 
two proper time ($\tau$, $\sigma$), but the particle's position can not be localized by each
of them individually. As the result, particle's spatial position $x_i$ is not localized 
at fixed time t. Therefore, different from Relativity, the physics of single particle is not localized 
by 4-dimensional space-time.

From Fig1 we see that at $t = x_0 =0 $, particle's spatial position can be
at $-x_n$ through the path: from $-\tau_n$ to $-x_n$; and can be at
$x_2$ through the path: first from $-\tau_n -> -\tau_2$, then from $-\tau_2 -> x_2$;
and can be at $x_1$ through the path: first from $-\tau_n->-\tau_1$, 
then from $-\tau_1 -> x_1$. 
At t=0 and $\frac{m_0\tau }{\hbar} = -2\pi$, we have
\begin{equation}
x_i = \frac{h}{p_i}
\label{wavelength}
\end{equation}
similarly at $x_i=0$ and  $\frac{m_0\tau }{\hbar} = 2\pi$, 
\begin{equation}
t = \frac{h}{E}
\label{period}
\end{equation}
If $\frac{m_0\tau }{\hbar}$ satisfies periodic condition, then equation(\ref{wavelength})
and (\ref{period}) become de Broglie equations, but $\tau$ is the proper time associate with time dimension
t, and we never observed periodic properties in classical physics for t, so we need to 
add one more time dimension and proper time in the next section.

 The Above equations illustrate the motion of single particle under two proper time with constant
energy and momentum. The single particle spreads out everywhere in space-time, i.e., with fixed
energy and momentum, the particle's position and time are uncertain.
That is, because at each fixed $\tau$, particle can move by $\sigma$ to another position. To localize
a particle, we need to make all $\sigma$ ``stay'' only at one position $(x_{i0},t)$. From
equation (\ref{Speed}), $v_i$ can not be zero, so world line $\sigma$ must be a infinitesimal 
loop around $(x_{i0},t_0)$ as shown in Fig3. But world line $\tau$ perpendicular to world line $\sigma$,
and the slope of $\tau$ is $\frac{u_i}{c}$; momentum $p_i = mu_i$. From Fig3, we see that because 
at each points on circle $\sigma$, particle can move perpendicular to $\sigma$ which create 
a world line $\tau$, and the slope of $\tau$ is from $-\infty$ to $\infty$, so the momentum becomes
$-\infty$ to $\infty$ . I.e. if we localized particle's position and time, the momentum and energy will become uncertain.

\begin{figure}[t]
\begin{center}
\leavevmode
\hbox{%
\epsfxsize=2.5in
\epsffile{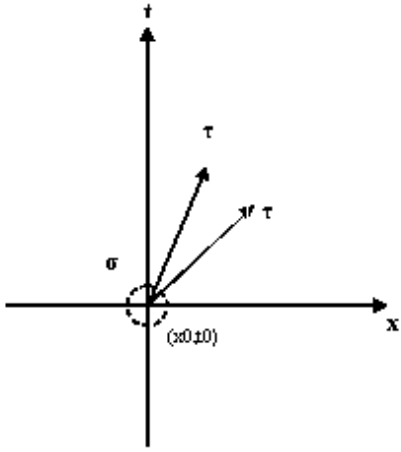}}
\caption{World lines $\sigma$ is a infinitesimal loop to fixed point ($x_0$, $t_0$).
$\tau$ perpendicular to loop $\sigma$, so $\tau$ can point to any direction, the
slope of $\tau$ is from  $-\infty$ to $\infty$ which means the momentum is
from $-\infty$ to $\infty$. }
\end{center}
\label{fig:Fig3}
\end{figure}

\section{Three proper time model -- Approach to quantum physics} \label{ThreeTime}

 Let's introduces new time coordinate $x_4$ and new proper time $\phi$. 
Fig4. draws time loop on a complex-plane.
\begin{figure}[t]
\begin{center}
\leavevmode
\hbox{%
\epsfxsize=2.5in
\epsffile{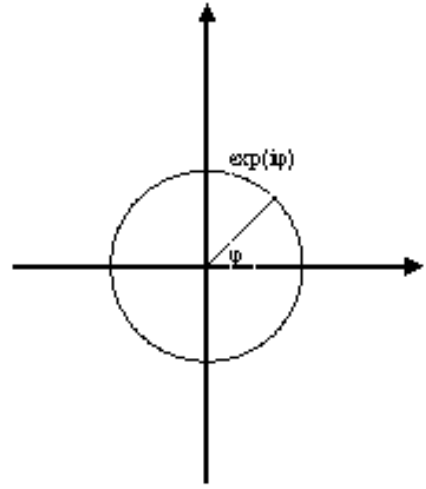}}
\caption{The loop of extra time dimension on complex plane }
\end{center}
\label{fig:Fig4}
\end{figure}
and $\phi$ is the angle from 0 to $2\pi$. The equation of loop is $e^{i\phi}$.
Let
\begin{equation}
\phi = \frac{m_0 \tau }{\hbar} 
\label{phi}
\end{equation}
i.e., particle moves around the loop of $\phi$ with angular velocity $\frac{m_0}{\hbar}$, 
uses equation (\ref{tau}), the equation of loop becomes:
\begin{equation}
e^{\frac{im_0 \tau }{\hbar} } = e^{\frac{i}{\hbar}(Et - p_{1}x_{1} - p_{2}x_{2} - p_{3}x_{3})}
\label{planeWave}
\end{equation}
It is wave function of quantum particle with fixed 4-momentum. 
In fact, in Fig1, at t=0, particle's positions
are any points from $-\lambda = \frac{h}{p_i}$ to 0 with $x_4$ from $-2\pi$ to 0; At x=0, particle
stays position (x=0) at time from 0 to $\frac{h}{E}$  with $x_4$ from 0 to $2\pi$. Therefore, Fig1. 
and Fig 2. illustrate
a plane wave. It is created by single particle's motion under
proper time $\tau$, $\sigma$, $\phi$. 

Assume the geometry of new time dimension $x_4$ is a loop with fixed radius 
(we can also assume the radius of loop is very small,
for example, the length of Planck constant h, to be convenience, here we choose radius =1 ), then 
\begin{equation}
x_4 = e^{i\phi}
\label{x4}
\end{equation}
On $x_i$ coordinate, when $x_4$ moves around a circle, $x_i$ moves from 0 to $\lambda$, this can be interpreted
as x oscillating along $x_4$ direction which is perpendicular to x-t plane. 

Let's go back to proper time $\sigma$, and let the time dimension which associates with $\sigma$ be $x_5$. Particle moves
along world line $\sigma$ with speed $\frac{c^2}{u}$ which is phase velocity of de Broglie wave. Similar to
$x_4$, assume $x_5$ is also a loop as illustrated in Fig4., we can build 1 to 1 relationship between loop $x_5$
and world line $\sigma$, let 
\begin{equation}
x_5 = e^{\frac{-i}{\hbar}(Et - p_{1}x_{1} - p_{2}x_{2} - p_{3}x_{3})}
\label{x_5}
\end{equation}

In real world, we only have knowledge of one dimensional time t. In experiment, we 
measure one time dimension by using ``clock''. We do not know how to synchronize each particle's 
2nd and 3rd time dimensions $x_4$ and $x_5$. When a particle of apparatus arrives at (x,t) with
2nd time dimension $x_{4a}$, the $x_4$ value of the particle to be measured can be any value on loop 
$e^{i\tau}$; But to ``meet'' the particle at location $X(x_i)$, the apparatu's particle
and the particle p must arrive at location $X(x_i)$ at the same three dimensional time, i.e.
$x_{4a} = x_{4p}$, and  $x_{5a} = x_{5p}$.
The possibility of $x_{4a} = x_{4p}$ is 
\begin{equation}
\psi = P_\phi = \frac{e^{i\tau}}{\int e^{i\tau}} 
\label{chance}
\end{equation}
Similarly, the possibility of $x_{5a} = x_{5p}$ is
\begin{equation}
P_\sigma = \frac{e^{-i\tau}}{\int e^{-i\tau}} = \psi^{*} 
\label{chance2}
\end{equation}
So the total possibility to find particle at (x,t) is 
\begin{equation}
P = P_\phi P_\sigma = \psi \psi^{*}
\label{possibility1}
\end{equation}

For the plane wave of single photon, proper time $\tau = 0$, equation (\ref{phi}) is no longer valid. Then proper times
$\sigma$ and $\phi$ are not related to $\tau$. Instead, we separate the motions of photon by three proper time.
1) Photon moves along world line $\tau$. 2) Photon has oscillation $E = E_0e^{-i\omega t + \lambda x}$ by $\sigma$
which perpendicular to $\tau$, where E is electric field. 3) Photon has oscillation 
$B = B_0e^{-i\omega t + \lambda x}$ by $\phi$
which perpendicular to $\tau$, where B is magnetic field. The possibility to find photon is
proportional to : 
\begin{equation}
S = E \times B
\end{equation}

In another paper \cite{chen2}, I proved that by choosing 6-dimensional space-time metric as
\begin{equation}
\left( \hat{g}_{AB} \right) = \left( \begin{array}{cc}
   g_{\alpha\beta} \; \; \; \; \; \;  \; \; \; \; \; \; \\
  \; \; \; \; \; \; \; \; \psi \; \; \; \; \; \; \; \; \\
   \; \; \; \;  \; \; \; \; \; \; \; \; \; \; -1 \\ 
   \end{array} \right) 
\label{6dMetric_0}
\end{equation}
where metric elements $g_{\alpha\beta}$ is 4-dimensional metric. We can derive Klein-Gordon equation directly
from Einstein field equation:
\begin{equation}
\hat{G}_{AB} = \kappa \hat{T}_{AB} \; \; \; ,
\label{5dEFE1}
\end{equation}
Under this metric, for spinless free particle, we have equation of world line $\tau$:
\begin{equation}
ds^2 = dx_{\alpha}dx^{\alpha} + e^{\frac{i}{\hbar}(p^{\alpha} x_{\alpha} - m_0 x_5)}dx_4 dx^4 - dx_5 dx^5
\label{metric1}
\end{equation}
Equation for world line $\sigma$
\begin{equation}
ds^2 = dx_{\alpha}dx^{\alpha} - dx_4 dx^4 + e^{\frac{-i}{\hbar}(p^{\alpha} x_{\alpha} - m_0 x_4)} dx_5 dx^5 
\end{equation}
In general, for spinless particle, equation (\ref{metric1}) becomes
\begin{equation}
ds^2 = dx_{\alpha}dx^{\alpha} + \psi^2 dx_4 dx^4 - dx_5 dx^5
\label{metric2}
\end{equation}
Put it into Einstein field quation, it satisfied wave equation \cite{chen2}:
\begin{equation}
\partial_{A} \partial^{A} \psi = 0
\label{wavefunction}
\end{equation}
where A = 0...5, $\psi$ is wave function. The wave equation (\ref{wavefunction}) is derived directly from Einstein field 
equation where Planck constant plays the same role as gravational constant\cite{chen2}.
It means that quantum phenomena can be understood as pure geometry effect of 6-dimensional space-time.

\section{Interpretation of quantum physics} \label{Interp}

Non-local property of single quantum particle is one of the most important reasons why quantum physics does not fit in 
classical physics theory. In quantum physics, a single particle can stay at different places at the same time.
In double slits interference experiment, if we try to use classical paths to describe particle's motion, the particle has to
pass both slits at the same time. In condense matter physics, a electron's spatial 
positions will be everywhere in lattice at any time; i.e. the electron's must be able
to stay in may spatial positions inside lattice at the same time; in the experiment about Bell's inequality, a single
particle must stay in two different spatial places even though the distance between those two places are ``far''. 
Those all conflict with our knowledge in classical physics(including Relativity). 
In classical physics with one dimensional time, a particle can not stay in more than one place at the same time.

Section \ref{TwoTime} demonstrates that by introducing multiple proper time, single particle's
motion shows non-local properties in classical physics. The particle can move to different places 
by extra proper time $\sigma$, in Fig1, at time t=0, particle's spatial positions are $x_1$, $x_2$, ... $x_n$.
In fact, section \ref{TwoTime} and \ref{ThreeTime} draw two different pictures:

1)Along world line $\tau$, free particle moves like classical particle 
with constant velocity with classical energy and momentum. 

2)World lines of $\sigma$ and $\phi$ are also straight lines. 
Relations between $\phi$ and 2nd time dimension $x_4$, 
$\sigma$ and third time dimension $x_5$ are $x_4 = e^{i\phi} = e^{\frac{im_0 \tau}{\hbar}}$, $x_5 = e^{i\sigma} =
e^{\frac{-im_0 \tau}{\hbar}}$, 
The equations come from the geometry of $x_4$ and $x_5$, which are loops in complex plane.

Put 1) and 2) together, then with equation it created plane wave function for single particle. 
Single particle's position is not unique under 1 dimension time t, but the position is unique under three dimensional
time (t, $x_4$, $x_5$). Because of the uniqueness,  
one particle can not contribute two energies, one electron can not contribute two electronic charges. 
In addition, at each point (t, $x_4$, $x_5$) of three dimensional time, the particle's energy, momentum
and charge are the same definition as in classical physics. 

To measure particle p at (x,t), the values $x_4$ and $x_5$ of apparatus's particle must by the same as the values
of p. That is because two things can only meet at the same time (here are three dimensional times). In section
\ref{ThreeTime} we have discussed that: because we do not have apparatus to measure 2nd and 3rd dimensional time,
we can not determine particle's location by our 4-dimensional apparatus. Instead, the particle's position is
statistical in 4-dimensional space-time description. The possibility of finding  
the particle is proportional to  $\psi \psi^{*}$ in equation (\ref{possibility1}). Equation (\ref{possibility1}) 
is derived by single free particle with constant velocity. In general cases, the map between $x_4$ and $x_5$ 
loops to 4-dimensional coordinates are not necessarily to be plane wave function. Instead, it could be the
combination of plane wave function with different frequencies and wave-lengths. For example, the case of Fig3.
Then the world lines of particle become multiple lines of world lines $\tau$ with different slopes. But on each
individual world line $\tau$, $x_4$ and $x_5$ are plane waves with the same frequency and wave-length. That means in general:
\begin{equation}
x_4 = \Psi =  \psi(\tau_0) + \psi(\tau_1) + ... \psi(\tau_n)
\end{equation}
Then
\begin{equation}
x_5 =  \psi^{*}(\tau_0) + \psi^{*}(\tau_1) + ... \psi^{*}(\tau_n) = \Psi^{*}
\end{equation}
So the possibility of finding particle is always $|\Psi|^2$.
When we found the particle, the particle's momentum, energy and other observables are defined along $\tau$, so
the average value of classical observable F is:
\begin{equation}
<F> = \int F |\psi|^2
\end{equation}

Fig1. and Fig2. show that when particle's momentum is fixed, particle's position is uncertain because particle can move
by $\sigma$. Fig3. shows that when particle's position is localized, particle's momentum becomes uncertain
because of the changing slopes of $\tau$ by world line $\sigma$, i.e. we can not find a world lines distribution of $\tau$ and 
$\sigma$ such that both position and momentum are constant. That is the reason we have uncertainty 
relationship for x and p in quantum physics.

Now let's consider the double slits interference experiment for particle. In x-y plane, let particle move along
x coordinate with y = 0, the double slits at x= d, two slit's coordinates are $S_1(d,y/2)$ 
and $S_2(d, -y/2)$, and screen at x = S. When particle reaches x=d, we know that even though the particle's
y component of velocity $u_y$ is zero: $u_y = 0$, particle still moves in y direction by proper time
$\sigma$, so at x=d, particle can move from S1 to S2 by $\sigma$, i.e. the particle passes both slits at 
the same time t. At $S_1$, the particle's $x_4$ and $x_5$ value are ($e^{i\phi}$, $e^{-i\theta}$), and at $S_2$ with values
($e^{i(\phi+\delta)}$, $e^{-i(\theta+\delta)})$, $\delta$ is a small number 
since the distance between $S_1$ and $S_2$ is small. After particle passes S1 and S2, the world line $\tau$ 
splits into two paths with world lines $\tau_1$, $\tau_2$. Let $x_{41}, x_{51}$ be value of $x_4$ and $x_5$ 
on path 1; $x_{42}, x_{52}$ be value of $x_4$ and $x_5$ on path 2.   
Suppose $\tau_1$ and $\tau_2$ meet at p where p is a point on screen. At p, we have 
\begin{eqnarray}
x_{41} = x_{42} \; \; \; \; 
x_{51} = x_{52}
\label{interference}
\end{eqnarray}
Let $\Delta L $= path(from S2 to p) - path(from S1 to p), then to get equation (\ref{interference}),
$\Delta L$ must satisfies:
\begin{equation}
\Delta L = (n + \frac{\delta}{2\pi}) \lambda
\label{interf1}
\end{equation}
where n is any integer and $\lambda$ is wave length, 
particle can not reach those points, which does not satisfy equation (\ref{interf1}); 
so we get interference pattern on the screen.

\section{Causality in three dimensional time} \label{Causality}

In real world, time has direction. Here proper time $\tau$ and proper time $\sigma$ have 
directions too.

\begin{figure}[t]
\begin{center}
\leavevmode
\hbox{%
\epsfxsize=3.0in
\epsffile{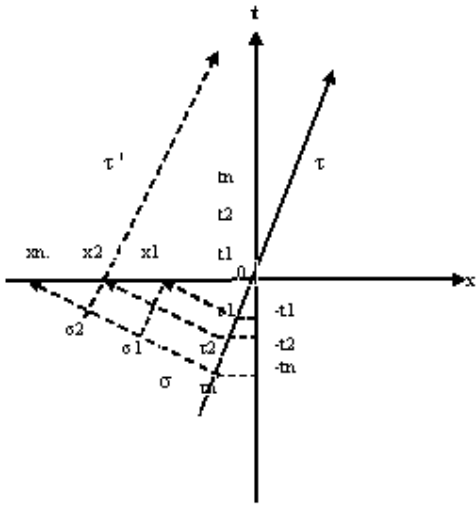}}
\caption{Current universal time is $t= -t_2$,
particle reaches $x_2$ at t=0, it is in future since $t=0 > -t_2$. When event 2 happens
at $x_2$, the particle interacts with other particles, so particle's world lines
is changed, its next movement will be based on new world line $\tau'$}
\end{center}
\label{fig:Fig5}
\end{figure}

From Fig1., Fig2. and Fig5., one can see that $\tau$ and $\sigma$ both move to positive direction
of t. Along world line $\tau$ when $\sigma$ is unchanged, if event 1 happens at $\tau_1$ and event 2 happens at 
$\tau_2$ and $\tau_1 > \tau_2$, then corresponding universal time $t_1 > t_2$, so, on world line 
$\tau$, the causality is preserved. Along world line $\sigma$ when $\tau$ is unchanged, if event 1 happens at 
$\sigma_1$ and event 2 happens at 
$\sigma_2$ and $\sigma_1 > \sigma_2$, then corresponding universal time $t_1 > t_2$, so on world line 
$\sigma$, the causality is also preserved. In general, event 1 happens at $(\tau_1, \sigma_1)$,
and event 2 happens at $(\tau_2, \sigma_2)$, if $\tau_1 > \tau_2$ and $\sigma_1 > \sigma_2$, then we 
have $t_1 > t_2$, causality is preserved. But what will happen when $\sigma_1 > \sigma_2$ and 
$\tau_1 < \tau_2$? On Fig5, if event 1 happens on $x_1$, event 2 happened on $x_2$, then on ``local static''
reference frame, event 1 happened after event 2 because $\tau_1 > \tau_2$, on universal time, event 1 happened 
at the same time as event 2 because both happened at t = 0, on world line $\sigma$, event 1 happened before
event 2 because $\sigma1 < \sigma2$, does it conflict with causality? 

Look at Fig5. Suppose universal time t at $t=-t_2$, particle reached $\tau_2$ along world line $\tau$;
then particle moves to $x_2$ at t=0 along world line $\sigma$, $\sigma = \sigma_2$, 
on ``local rest'' reference frame (world line $\tau$) which is still at $\tau = \tau_2$, the particle goes
to future because $t=0 > t_2$. If at $x_2$, particle does not have any interaction with other particles, then
the particle can't
``see'' anything in future, when particle moves back to $\tau = \tau_1, \sigma = 0$, there is nothing related
to causality, no event occurred. But if event 2 happened at $x_2$; i.e, particle interacts with other particle,
then the particle's physical state is changed by interaction (Remember: We can not measure a particle without 
affect its physical status). Suppose the particle's momentum has a small changes: $\delta p$, then the particle's
next move will start on a new world line $\tau '$, the particle can not go back to original world line
$\tau$, this phenomenon is corresponding to wave-packet collapse in quantum physics, so any event happened on this particle
after event 2 will be on time $t > 0$, the causality is still preserved. Oriented $\tau$ and
$\sigma$ and wave-packet collapse are key factors to keep causality preserved. 

All physical reference frames still move along world line $\tau$ with speed $u < c$, so causality is preserved
in any reference frame. Although particle's speed $v > c$ along world line $\sigma$, we can not observe or
measure this speed because we can not determine particle's position without affecting particle's velocity (momentum).
From all above, we see that three dimensional time contain the basic properties of quantum physics, one
can understand that three dimensional time will not conflict with causality law unless
quantum physics itself conflicting causality law.
If two identical particles are correlated each other, and we separate the wave to two parts with certain
distance, if we affect one part of wave, then the other part on the other place will be affected on the same time
(i.e. the information passed without time change). This is the well known Bell's inequality. There are already
many papers discussing about causality law in this phenominon \cite{holland}.

\section{Some discussinons} \label{Discusssion}

First, it is interesting to see the relation between three dimensional time and String theory. Actually
Fig1. looks like a world sheet in String theory. If we let $\sigma$ be space dimension 
instead of proper time, it turns to bosonic String theory. There are two major differences between three
dimensional time and String theory.

1) In String theory, the motion in $\sigma$ direction is compacted. 
It can not be very large since we never see
extra space dimensions in real world. In three dimensional time, the distance traveled by world line $\sigma$ 
can be very large, this is a very important property which
provides non-local properties of three proper time physics.

2) Three dimensional time has different statistical results from String theory. Because of the special character
of time (which different from space), it demonstrates the same statistical results as quantum physics.

But we still can use some results of String theory. Considering two proper time case, put Lagrangian
\begin{equation}
L = - \frac{1}{2} m [(\dot x \cdot x')^2 - (\dot x \cdot \dot x)(x' \cdot x')]^{1/2}
\label{Lagrangian}
\end{equation}
Where $\dot x = \frac{dx}{d\tau}$; $x'= \frac{dx}{d\phi}$.
The Lagrangian above is the same as the Lagrangian in string theory \cite{Collins1}. 
The classical
equation of motion is
\begin{equation}
\frac{\partial}{\partial \tau} (\frac{\delta L}{\delta \dot x_\alpha} + 
\frac{\partial}{\partial \phi} (\frac{\delta L}{\delta \dot x_\alpha})) = 0
\label{motionEq}
\end{equation}
Add constraints \cite{Collins1}:
\begin{eqnarray}
\dot x \cdot x' = 0 \;, \;\;\; \dot x \cdot x + x' \cdot x' = 0
\label{Constraints1} 
\end{eqnarray}
The equation of motion (\ref{motionEq}) becomes wave equation:
\begin{equation}
\ddot x_\alpha = x_{\alpha} ''
\label{waveEq}
\end{equation}
The above result is the same as bosonic string theory \cite{Collins1}. For free particle
with constant momentum, we choose solution:
\begin{eqnarray}
x_\alpha(\tau , \phi ) = e^{-\frac{im_0}{\hbar}(\tau - \phi)} 
\label{MotionEq3}
\end{eqnarray}
Since $x_\alpha(\tau , \phi )$ must be real number, the above equations have solutions 
only when
\begin{equation}
\phi =  \tau = \frac{1}{m_0}(Et - p_{1}x_{1} - p_{2}x_{2} - p_{3}x_{3})
\label{MotionEq5}
\end{equation}
We see that equation(\ref{MotionEq5}) corresponding to proper time $\phi$.

Second, there are possible some interesting relations between three dimensional time and quantum field 
theory. Feynman \cite{Feynman} interpreted negative energy state of particle as: negative energy state
represents the particle moving to negative time direction. It is hard to understand or illustrate this in 
1 dimensional time theory, but it can be often seen in three dimensional time: looking at Fig5. particle goes
to $x_2$ (future) at t=0, then go back to $t_1(\tau_1, \sigma = 0)$. 
In addition, in section \ref{TwoTime}, Fig3, when
$\sigma$ accrossing $X_i$ coordinate, momentum becomes infinite, that is because the particle moves from one
location to another location without changing universal time t -- the particle moves by $\sigma$ only. That is to say 
We get infinity momentum because we only use space and first time dimension t to calculate momentum. If
we uses $\sigma$, the infinity will be gone. It is
possible to use this to deal with the infinities in quantum field theory in future. 

Third, this paper is only dealing with ``basic'' quantum physics. I.e., it is only discussing spinless particle. 
I believe that spin is coming from the motion of extra time dimension. To discuss the particle with 
integer spin and half-integer spin, we have to find a way to explain the results of 
Bose-Einstein statistics and Fermi-Dirac statistics. We will discuss that in another paper. 

I used three dimensional time to interpret quantum physics in two other papers before \cite{chen1} \cite{chen2}.
This paper provides more details and clear pictures for three dimensional time theory.

\acknowledgments

The author acknowledges professor Yujie Yao in Jilin University for her encouragement
in 1991 when author told her the original idea about interpreting quantum physics by
three dimensional time.

\end{document}